\documentclass[preprint,review, 2pt]{elsarticle}

\usepackage{amssymb}
\usepackage{multirow}
\usepackage{url}

\journal{Nuclear Physics B}

\begin{document}

\begin{frontmatter}



\title{Human and AI Trust: Trust Attitude Measurement Instrument} 





\author[inst1]{Retno Larasati}

\affiliation[inst1]{organization={Knowledge Media Institute, The Open University UK},
            addressline={Walton Hall}, 
            city={Milton Keynes},
            country={United Kingdom}}

\author[inst1]{Anna De Liddo}
\author[inst1]{Enrico Motta}

\begin{abstract}
With the current progress of Artificial Intelligence (AI) technology and its increasingly broader applications, trust is seen as a required criterion for AI usage, acceptance, and deployment. A robust measurement instrument is essential to correctly evaluate trust from a human-centered perspective. This paper describes the development and validation process of a trust measure instrument, which follows psychometric principles, and consists of a 16-items trust scale. The instrument was built explicitly for research in human-AI interaction to measure trust attitudes towards AI systems from layperson (non-expert) perspective. The use-case we used to develop the scale was in the context of AI medical support systems (specifically cancer/health prediction). The scale development (Measurement Item Development) and validation (Measurement Item Evaluation) involved six research stages: item development, item evaluation, survey administration, test of dimensionality, test of reliability, and test of validity. The results of the six-stages evaluation show that the proposed trust measurement instrument is empirically reliable and valid for systematically measuring and comparing non-experts' trust in AI Medical Support Systems.
\end{abstract}



\begin{keyword}
Human-AI Trust \sep Human-AI Interaction \sep Trust Factors \sep Trust Measurement
\end{keyword}

\end{frontmatter}


\section{Introduction}

Nowadays, the study of trust in Artificial Intelligence (AI) is of great concern in computer science and cognitive systems engineering and is also becoming a hot discussion topic in the media. The call for trustworthy AI was made by formal national institutions around the world (Europe \cite{commissi36:online}, USA \cite{DIBAIPRI32:online}, China \cite{TheEthic13:online}). 
Trustworthiness in AI systems is increasingly becoming an ethical and societal need. Trust is a crucial factor in all kinds of relationships, such as human-human social interactions or human-AI system interactions. Trust is humans’ primary reason for acceptance \cite{gefen2003trust}. A 2018 survey conducted by Intel shows that 36\% of patients lack trust in AI  and identify trust as a key barrier to AI adoption \cite{USHealth39:online}. People in general are sceptical about AI, and in those instances when AI did something wrong, such as, the Google Photos algorithm classifying black people as gorillas \footnote{https://mashable.com/2015/07/01/google-photos-black-people-gorillas/}, or the Microsoft chatbot that turned racist in a day \footnote{https://www.theverge.com/2016/3/24/11297050/tay-microsoft-chatbot-racist}, the general public could not understand why the AI did it, and were therefore left only with a sense of distrust toward such systems. This process will only worsen if the public keeps receiving news about the harm of AI. For example, there was an uproar in 2017 over the UK's National Health Service (NHS) allegedly illegally handing 1.6 million patient records to Google's DeepMind, as part of a trial \cite{powles2017google}. This news sparked conversations about privacy and ethics. In 2018, a government-backed AI healthcare application, Babylon, also received criticism for the inaccuracies in diagnosis, \cite{ThisHeal50online} which brought the medical regulator, the Medicines and Healthcare products Regulatory Agency (MHRA), into the spotlight. These controversies only add to the reported general unwillingness of people to engage with AI when it gets to their healthcare needs \cite{Surveyre41:online}.

However, research also shows that people who are already AI users tend to easily take algorithmic outputs as accurate and valid and even prefer an algorithmic decision to human advice \cite{logg2019algorithm}. People that are more accustomed to AI-facilitated processes have been shown to over-trust AI recommendations, even when the AI systems have been proved to malfunction or when the use of the system caused harm. As AI is increasingly embedded in all sorts of largely adopted systems, research evidence indicates that users tend to over-trust and continue to rely on a system even when it malfunctions \cite{cohen2017ismp}. This phenomenon is known as automation bias, which occurs when people tend to over-trust and accept system outputs ‘as a heuristic replacement of vigilant information seeking and processing' \cite{goddard2011automation,mosier2018human}. People often neglect automation bias and tend to trust a system when they think the answer came from an algorithm rather than another person \cite{logg2019algorithm}. This misplaced trust (distrust or over-trust) has motivated research on trustworthy AI, whether via developing new forms of explainable AI or AI transparency \cite{wachter2017counterfactual,poursabzi2018manipulating,abdul2018trends,zhang2020effect}. 

A huge challenge into advancing research in this crucial research field is the issue of comparability. Currently, there is no general trust measurement or evaluation method for research in AI trust. In evaluating trust, literature measures users' confidence \cite{salomons2018humans,bridgwater2020examining,antifakos2005towards}, reliance \cite{yuksel2017brains,feng2019can}, and also straightforward trust rating \cite{bridgwater2020examining,yin2019understanding,bussone2015role} amongst other evaluation methods. Aside from the context specific nature of trust \cite{li2019no}, the difference in evaluation approach and how trust is measured stemmed from variations in trust definition, which lead to different trust metrics, and therefore prevents from meaningful scientific comparisons. Another important point to note is the broad definition of Artificial Intelligence (AI). To put it simply, Artificial Intelligence is artificially constructed intelligence, it ranges from prediction to recommendation system; to physical materialization, such as, robots and automated machines. This variety results in different trust measurements which rely on questionnaires derived from research in the human-robot trust \footnote{The "-" symbol is used as an indicator of the trustor and trustee in the interaction between both. Human-robot indicates interaction between human as a trustor and robot as a trustee in the interaction.} \cite{schaefer2013perception}, human-automation trust \cite{jian2000foundations,muir2002operators,merritt2011affective}, and human-technology trust \cite{mcknight2002impact}, which have been used interchangeably between sub-fields. Combination of existing questionnaires was also implemented to achieve usable measurement tool \cite{ghai2021explainable,yu2020keeping,cheng2019explaining}. However, most of the research has not conducted any validation test \cite{vereschak2021evaluate}, or only conducted one test (reliability\cite{ghai2021explainable} and predictive validation \cite{cheng2019explaining}), to their adapted questionnaires. This lack of measurement validation raises concerns and can undermine the validity of research findings achieved using said measurements. Moreover, valid measurement instruments play a significant part in the progress of trustworthy AI design, development, and research; how can we design trustworthy AI systems if we cannot measure the effect of our design choices in a reliable and comparable way? 

In this study, we developed and validated a trust measure instrument following the psychometric principles, methodological concepts, and techniques in scale development and validation research \cite{boateng2018best, hinkin1998brief, murphy1988psychological, raykov1997scale, american1999standards, campbell1959convergent, nunnally1994psychometric, raykov2011introduction, devellis2021scale}. The instrument is specifically built for research in human-AI interaction, to measure trust attitude towards AI systems, from a layperson (non-expert) perspective. The use-case we used was in the AI medical support system context (cancer/health prediction). The development (Measurement Item Development) and validation (Measurement Item Evaluation) involved six stages, which are: item development (Section 4), item evaluation (Section 5), survey administration (Section 6), test of dimensionality (Section 7), test of reliability (Section 8), and test of validity (Section 9). 

Our work provides three main contributions to the field. Firstly, we demonstrate a methodological approach to develop and thoroughly evaluate a trust measurement for human-AI interaction. Secondly, we propose a trust measurement instrument to evaluate trust in human-AI interaction for laypeople, which future researchers can use and adapt to evaluate their AI systems. Finally, our work contributes to the ongoing conversation regarding trust and trust measurement in human-AI interaction.

\section{Background and related work}
\subsection{Trust Concepts}
Mayer et al. conceptualised trust as a willingness to be vulnerable based on the expectation that another party (the trustee) will perform certain actions that are important to the trust giver (trustor), regardless of the ability to monitor or control the trustee \cite{mayer1995integrative}. Although the context of Mayer et al.'s trust concept is human-human trust in organisations, this definition was widely applied and adapted in the context of human-technology trust, such as, trust in automation \cite{lee2004trust}\cite{schaefer2016meta}, trust in information systems \cite{mcknight2011trust}\cite{li2008we}, and trust in robots\cite{oleson2011antecedents}\cite{hancock2011meta}. This definition of trust, and its adaptations, emphasise the components of competence and vulnerability within it and consider them as factors that can influence trust. 

Mayer et al. noted the distinction between trust, as in the factors that influence trust (trust factors), with trust-related behaviour, irrespective of the relationship between these two. Trust as an attitude does not always translate into trust-related behaviours, such as, dependence and \cite{meyer2013trust}, and should be measured separately. In contrast, although the concepts of trust and trust factors are easily distinguished, the measurement aspects are quite connected. Since trust is regarded as an attitude, which is a "psychological construct, a mental and emotional entity attached to or characterising a person" \cite{perloff1993dynamics}, it is said to be externally non-observable \cite{mayer1995integrative,xie2019robot}. Psychological constructs are determined by psychological factors and can therefore be measured using self-reports, attitude scales, or questionnaires, for example, the Likert Scale \cite{likert1932technique}. The Likert scale is one of the scales that has been widely used and supported by the attitude measurement literature \cite{nunnally1994psychometric,biasutti2017validity,hair2019development}. This study focuses on measuring trust as an attitude through the means of trust factors using Likert scales. 

\subsection{Measuring Human Trust in AI Systems}
In general, there is a considerable trust measurement literature, be it behavioural trust (trust-related behaviour) or attitudinal trust (trust as an attitude) \footnote{Since we have established that behavioural trust and attitudinal trust are different, and therefore measured differently, the discussion below covers only the attitudinal trust-related literature}. Several disciplines, such as psychology \cite{rempel1985trust} and management \cite{mayer1995integrative}, have been looking at human trust in technology. In particular, much work has been done investigating trust in human-automation interaction \cite{lee2004trust,schaefer2016meta,lee1994trust,lewandowsky2000dynamics,moray2000adaptive} and human other technologies interaction \cite{mcknight2002impact,li2008we}. However, only some of these studies have included measurement scales \cite{chien2018effect, jian2000foundations, dzindolet2003role, schaefer2013perception}. Several trust measurement scales have become recurring trust scales used in human-AI research. One such scale, Jian et al.'s \cite{jian2000foundations}, is reported to be the most cited trust scale in human factors research. 

A closer look at the existing measurement scales show that some of these trust measurements are very specific to certain applications. For example, the scale developed by Schaefer \cite{schaefer2013perception} refers specifically to the context of human dependence on robots in team settings, with one of the questions: "Does the robot act as part of the team?". Dzindolet et al. \cite{dzindolet2003role} developed a trust measurement scale for AI systems, with one of the questions asked being: "How many mistakes do you think you will make over 200 trials?". These questions are highly specific to the task, to their application context, and to their user's expertise. More general scales were also developed \cite{jian2000foundations,madsen2000measuring,mcknight2011trust}. While the scale by Jian et al. was developed for human-automated systems trust \cite{jian2000foundations}, the survey questions comprising the scale are very generic, which can be one of the main reasons for its re-usability. The scale dimensions and items were developed through elicitation of trust definitions, followed by cluster and factor analyses. Jian et al.'s proposed scale contains 12 items, with seven items to measure trust and five items to measure distrust. A similarity matrix was used to prove inter-rater reliability. However, no validity for the scale was established. As mentioned previously, most of the research which utilised trust measurement instrument has not conducted any validation test \cite{vereschak2021evaluate}, or only conducted one validation test \cite{ghai2021explainable,cheng2019explaining}, which is not adequate. Measurement instrument or measurement scale should demonstrate internal consistency, and different types of validity: content validity, construct validity, and criterion-related validity, to be sufficient \cite{american1999standards}. 

Madsen and Gregor \cite{madsen2000measuring} developed a more generalised measurement for human-computer trust, and tested the measurement reliability and validity. The dimensions used in this measurement were common factors that influence trust. The factors were constructed using the Nominal Group Technique, and then compared to constructs from previous trust research. Through the scale validation process, high internal consistency (Cronbach's alpha > 0.94) and construct validity were established, with poor criterion-related validity. The final scale proposed by Madsen and Gregor consists of five main trust factors: such as, perceived reliability, perceived technical competence, perceived understanding, confidence, and personal attachment, with five questions proposed for each dimension. McKnight developed another trust measurement instrument to capture the trust relationship between users and specific technologies \cite{mcknight2011trust}. This scale was developed based on an understanding of trust in the broader context of society and previous research on human-human trust. After conducting several trust-related studies with different information systems, and by covering a large literature, including trust in humans, McKnight defined trust as a construct consisting of three components: propensity to trust, institution-based trust, and trust in specific technologies. The scale proposed by McKnight has eight main dimensions: perceived reliability, perceived functionality, perceived usefulness, situational normality, structural assurance, confidence, and trusting attitude. Three to four questions are proposed to measure each dimension. These measurement instruments show good reliability (Cronbach's alpha > 0.9), construct validity, and criterion-related validity. 

To date, however, there has been no developed trust measurement instrument intended specifically for AI non-expert users/laypeople. Although the more general trust measures described above have been used for laypeople, appropriate modifications to fit the context of our study are still required. Therefore, in the next section, we describe the steps we followed to achieve a suitable yet generalisable measurement instrument for non-expert users' trust in AI medical systems. 

\section{Measurement Instrument Development Process}
\begin{figure}[h]
  \centering
  \includegraphics[width=15cm]{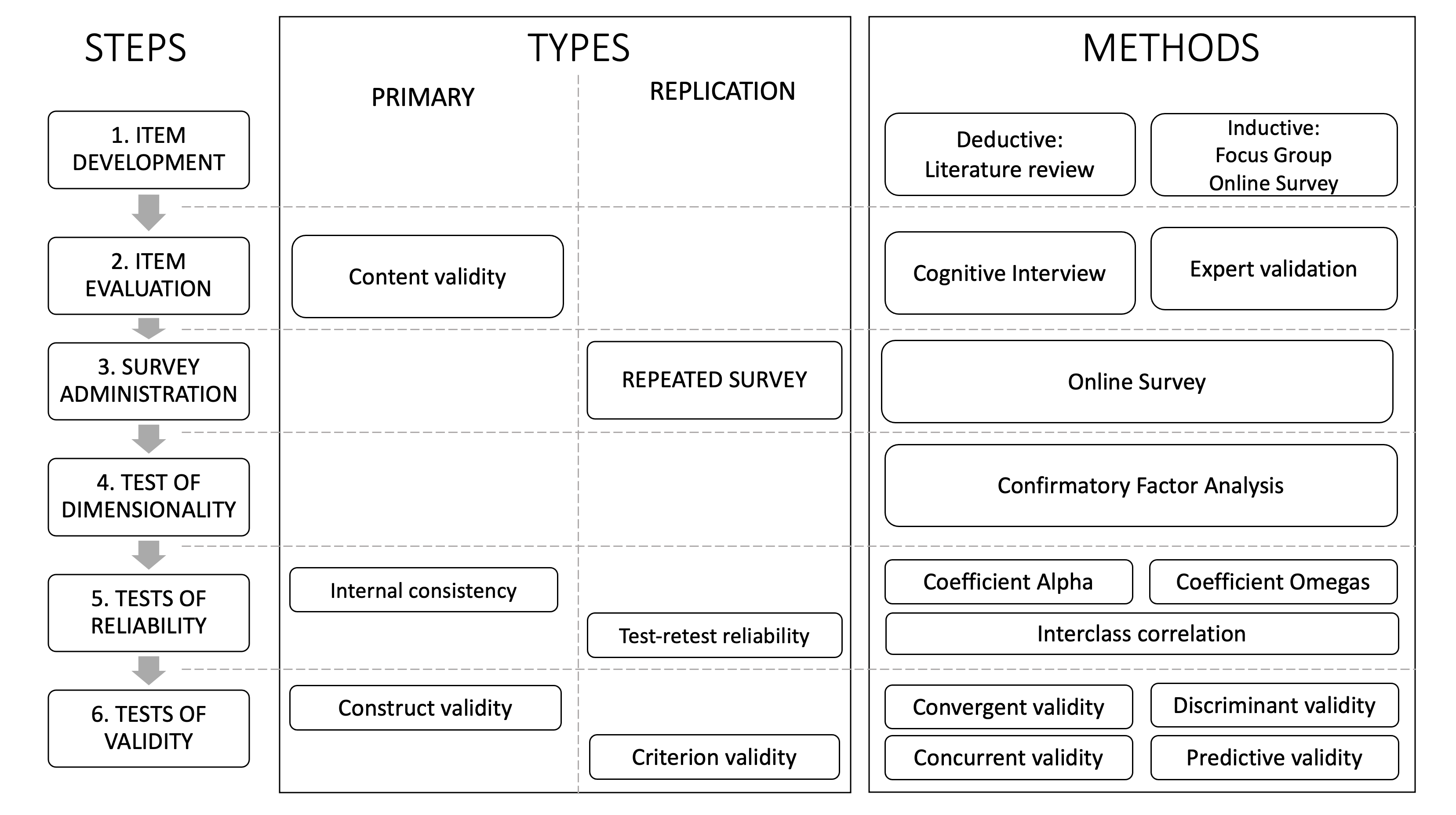}
  \caption{Six Steps Process To Develop a Sound Human Trust Measurement Instrument. }
  \label{fig:truins}
\end{figure}

To develop a sound human trust measurement instrument, we followed recommendations by previous research in psychometric \cite{boateng2018best,hinkin1998brief}. Six key research stages (Fig \ref{fig:truins}) were carried out to develop the items, and thoroughly evaluate them (through the assessment of each of the item individually, the overall scale, and the possible correlation between items). We finally carried out validity and reliability tests. According to the Standards for Educational and Psychological Testing, a guideline approved by American Psychological Association (APA), an appropriate operational definition of the construct a measure aims to represent should include a demonstration of content validity, criterion-related validity, and internal consistency \cite{american1999standards}.  The complete steps of the method and analysis we carried out is shown in Fig \ref{fig:truins}. The detailed description of each steps will be described in the next sections. 

\subsubsection*{1. Item Development}
The first step to developing a measurement instrument is to determine the \textit{measurement domain} that will be used to identify \textit{measurement items}. A measurement domain (or sometimes referred to as a measurement construct) is the concept or attribute which is the measurement target. In this study, the domain is a set of trust factors, as trust (the target) will be measured using various factors (attributes) that influence human trust in AI systems. The trust factors and the relevant items were generated using deductive and inductive approaches. As an example, Perceived Understandability is one of the trust factors with one of the relevant items is a statement "I know what will happen the next time I use the system because I understand how it behave".
The deductive approach requires a literature review to develop a domain definition with a theoretical foundation. Trust factors proposed in the literature were assessed and selected based on context relevance. The inductive approach requires exploratory research to develop items from dimensions that may not be easily identified in the conceptual basis. 
A mixed methods study comprising an online survey and a focus group were conducted to help explore other dimensions that may not have been considered previously. The study also helped contextualise and revise the items in the context of a concrete healthcare scenario. 
It should be noted that, while the measurement instrument was developed to be as general as possible to enable future use in AI system interaction evaluation, the use-case we used was an AI medical support system (cancer/health prediction).

\subsubsection*{2. Item Evaluation}
Once the measurement domains had been defined and the measurement items have been developed, we carried out an evaluation process. In the first evaluation process, the initial set of items was reviewed by experts. The measurement domains (trust factors) were presented and the experts were asked to provide a review and rating for each measurement item. The expert validation resulted in a number of items being reconstructed or being removed from the measurement instrument. The revised items were then further evaluated by the target population, in this case the general public, through cognitive interviews. The cognitive interview assessed how the target population understood the measurement domain and the mental processes behind the answers given from the measurement instrument. 

\subsubsection*{3. Survey Administration}
At this stage, the measurement items have been evaluated and revised. We then administered the survey as the main measurement instrument test. The sample size of the survey was carefully considered. The survey questions were presented to the participants after a description of random AI system. The measurement items were designed to quantitatively measure the trust factor, and we used a 7-point Likert scale for this survey. As a replication process, the survey was administered in a repeated manner. Replication was deemed necessary to increase the generalisability of the measurement instrument \cite{hinkin1998brief}. 

\subsubsection*{4. Dimensionality Test}
Dimensionality testing is the first part of measurement instrument evaluation. It evaluates the hypothesised factor or factor structure, and can be performed using confirmatory factor analysis, bifactor modelling, or measurement invariance. Confirmatory Factor Analysis (CFA) was conducted to quantitatively assess the measurement dimensions, thereby confirming that thorough analyses have been conducted to develop the measurement instrument. 

\subsubsection*{5. Reliability Test}
Reliability is the degree of consistency shown when a measurement is repeated under the same conditions. Different ways to assess reliability are: inter-temporal reliability (test-retest reliability) and inter-item reliability (internal consistency)\cite{cook2006current}. Internal consistency was assessed with alpha and omega coefficients, and stability across time was assessed with test-retest reliability. 

\subsubsection*{6. Validity Test}
Validity is the extent to which "evidence and theory support the interpretation of test scores required by the proposed use of the test" \cite{american1999standards}. In this step, we assessed the Construct Validity and Criterion Validity of the instrument.  Construct Validity is "the extent to which an instrument assesses a construct of interest and is linked to evidence measuring other constructs in that domain and measures specific real-world criteria" \cite{raykov2011introduction} and Criterion Validity is "the extent to which there is a relationship between a given test score and performance on another measure of special relevance, usually referred to as a criterion" \cite{raykov2011introduction}. 

\section{Item Development}
\subsection{Methods}
\subsubsection{Deductive Method}
The literature on human-computer trust was reviewed as the first step in the item development step. It is important to identify what we wanted to measure and what type of measurement instrument we wanted to develop. Next, we should identify whether validated measurement instruments are already available before developing new ones. From a review of the literature in the relevant field, no available instruments were identified to measure trust in the context of human-AI from the perspective of non-expert user/laypeople. 

We decided to develop a quantitative measurement, in the form of a survey, to measure human trust in AI systems based on factors that may influence human trust. As the construct of interest has been determined, we collected the literature related to trust from various fields, such as, Human-Computer Interaction, Information System, and Human-Factors. Indeed, much of the relevant literature cited and informed this research are not necessarily in the field of AI for the general public; therefore, an early user study to contextualise, extend and revise the theoretical framework of human-AI trust needs to be conducted. Hence, an inductive approach was also undertaken to discover factors that may have been missed from the literature review (deductive step).

\subsubsection{Inductive Method}
A mixed method approach of online surveys and focus groups was selected to gather people's general opinions. We chose online surveys because online surveys can reach a wide population of individuals with various characteristics, which better matches our target population of non-expert users \cite{easterbrook2008selecting}. We chose to use online surveys, rather than paper surveys, and certain crowd participation systems (such as Mechanical Turk) to provide access to a broader group, a larger number of potential participants, who could be reached in a timely and efficient manner" \cite{wright2005researching}. An initial version of the measurement instrument was developed to conduct the online survey using the initial domains and items. 

At the same time, since the answers from online surveys are usually quite short, we conducted additional data collection with focus groups to improve the depth of analysis. Focus groups are an effective way to collect rich data due to the nature of the interaction between participants \cite{mansell2004learning}. For example, unlike in the online survey, we explicitly asked participants what they thought the main challenges and opportunities were in using AI systems in healthcare, which helped improve our understanding of the underlined patterns identified in the online survey. To help contextualise the issues to healthcare scenarios, we provided participants in the online survey and focus groups with dramatised vignettes to read and reflect on. We aimed to evoke and elicit participant insights and chose the vignette technique as a method of enquiry. The vignette technique is a method that can elicit perceptions, opinions, beliefs, and attitudes from responses or comments to stories describing scenarios and situations that may be fictitious or adapted from real events" \cite{barter1999use}. According to Finch, a vignette is "a short story about a hypothetical character in specific circumstances, to which situation the interviewee is invited to respond \cite{finch1987vignette}. Vignettes are particularly appropriate for obtaining feedback from users regarding the implications of unrealised future possibilities. Therefore, this method is well suited for our research, where we need to elicit people's opinions on the public use of AI-assisted healthcare for early cancer diagnosis that does not yet exist.

We wrote a vignette depicting a fictitious scenario where AI-assisted health assessment is available and accessible to everyone for early cancer diagnosis. Although the vignette was fictional, the story still had to seem plausible and real to \cite{barter1999use} participants and had to be built around actual experiences. Our vignette was based on real experiences shared by breast cancer survivors on the breast health centre website \footnote{http://www.breastlink.com/}. The dramatisation of this vignette was designed, with the aim to stretch participants' thinking towards opposing views, extreme scenarios, and promote reflection on contentious actions and unforeseen consequences. 

For the online survey, questions were presented using a Google Form formed as long text answers. The online survey was published through Amazon Mechanical Turk. Our target was 80 participants, with 40 participants from the general public and 40 participants from workers in the healthcare field. We placed "master worker" as the selection criteria for participation and also placed one check-in question within the survey to validate if participants read the vignettes carefully. The Mechanical Turk task lasted for a week, and in the end, we had 53 participants. 

For the focus group, eight participants were recruited through the forum and asked to attend a face-to-face meeting on the Open University campus. Participants were Open University students and staff. The focus groups ran for one hour and were audio-recorded. The recordings of the focus group discussions were then transcribed, and any personal information was removed from the transcripts. 

Both the focus group transcripts and the online survey open-ended responses were analysed using a grounded theory approach. Several studies on healthcare systems and technologies have successfully applied the grounded theory method \cite{thom1997fotlg,winkelman2005patient}. Grounded theory is an appropriate approach to use in inductive methods because it allows the generation of new statements, hypotheses, or relationships. 

\subsection{Results: Trust Factors and Items Developed}
\subsubsection{Deductive Method}
As mentioned previously, the aim of this study is to develop a scale to measure trust attitude towards an AI system based on trust factors. We have the reviewed literature on trust and identified factors that could affect trust. 

Research on trust in human-automation systems' interaction, suggested that trust consist of human-related trust, environment-related trust, and learned trust \cite{lee2004trust}\cite{hoff2015trust}\cite{schaefer2016meta}. Human-related trust is, as the name suggest, related to the human trustee, such as individual personalities, backgrounds, and capabilities. As the name suggest, environment-related trust is related to the environment or situation of the task and the system. Learned trust relates to the system itself, such as its behavior, reliability, transparency, and performance. Other research also proposed similar form of trust in different trust context, composed by similar concepts with different names. For example, in human-robot context, trust is composed of human factors, environment factors, and robot factors\cite{oleson2011antecedents,hancock2011meta}. In the context of information systems, trust consists of basic personality trust, basic institutional trust, and basic system trust. \cite{mcknight2011trust,li2008we}. To put it simply, a person trust towards an object (person, robot, AI) is built from their personality/attributes/characteristics, the environment surrounding the person and the interaction, with the object and the attributes/characteristics of the object.  

A different perspective by Morrow et al theorised trust under two bases: cognitive and affective \cite{morrow2004cognitive}. Cognitive base trust is trust resulted from a pattern of careful and rational thinking, while, affective base trust is trust that results from feelings, instincts and intuition. Moreover, this theory was included in the literature on human-computer trust \cite{madsen2000measuring} and in human-automation \cite{schaefer2016meta} under the human-related factors affecting trust. The link between human-related factors of trust and categorisation of cognitive-affective base trust, shows that trust is a multidimensional concept and can be categorised in different ways. An inter-disciplinary exploration on trust theory concluded that trust concepts are actually similar, overlapping, and sometimes only divided by different jargon \cite{rousseau1998not}. Therefore, we looked at the literature on trust factors while not overly considering discipline-specific categorisations.

Understanding how the system works is part of cognitive factors in trust in automation \cite{schaefer2016meta,lee2004trust}. Not only for human-automation, human trust is often dependent upon such an understanding of a system/machine in general \cite{muir1987trust}. Empirical evidence has shown that understandability affect users' trust and confidence \cite{sinha2002role}\cite{herlocker2000explaining}. Cognitive compatibility, or understandability, was also found to be a trust factor in the study of trust in medical assistance devices \cite{hengstler2016applied}. 

Trust is dynamic, and can change over time as a result of experience with the system \cite{muir1987trust}. Research found that trust is typically built up in a gradual manner \cite{gefen2000commerce,mcknight1998initial} , and when a system offers consistent experience the system is seen to be "reliable" or "predictable".  In socio-psychology, reliability is seen as a factor of trust \cite{rempel1985trust,johnson1982measurement}. Similarly, reliability is considered as one promoting factor of trust in automation \cite{lee2004trust,schaefer2016meta,hoff2015trust}. At first, a person will judge the predictability of a system by assessing the consistency of its behaviours. The more consistent the system is, the more predictable it will appear to be. Trust will be directly related to the stability of the system, or system reliability. 

At the end of repeated positive experiences, a person may develop "faith" towards a system. Even when people cannot know that a system will continue to be reliable in the future, they need faith to continue to depend on the system in the future. Faith means emotional security or confidence despite potential future uncertainty, and is considered as a factor of trust \cite{rempel1985trust} which is driven by the affective/emotive side of a person \cite{johnson1982measurement}. Faith is also seen as a factor of trust in human-automation and human-computer context \cite{schaefer2016meta,madsen2000measuring}.

On the one hand, a positive experience could affect cognitive base trust and, over time, forms affective base trust. On the other hand, a negative experience could also affect trust judgement. People who initially trusted a system could turn to distrust when they encounter system faults or errors\cite{dzindolet2003role}. Since errors and system faults indicate the lack of system competency, lack of trust is then expected because competency is considered one of the antecedents of trust \cite{mayer1995integrative}. The display of system technical ability/competence, in the form of confidence level, also affects the user's trust in automatic notification devices \cite{antifakos2005towards}. Technical ability to correctly perform its tasks is frequently regarded as one of the most important factors for human-machine trust \cite{muir1987trust,madsen2000measuring} and human-automation trust \cite{hoff2015trust,schaefer2016meta,lee2004trust}.

Other than competency, benevolence is also seen as a trust antecedent in human-human trust \cite{mayer1995integrative,rempel1985trust}. Benevolence means that the trustee is believed to \textit{want} to do "good" to the trustor. In human-automation and human-AI trust research, benevolence is seen as a system's purpose \cite{lee2004trust,schaefer2016meta}. However, since technology has no agency (want) nor intention (purpose) \cite{mcknight2011trust}, we could argue that benevolence can also be seen as helpfulness in the context of human-technology trust. Therefore, helpfulness means that trustee is believed to \textit{be able} to do good to the trustor, and is included as a the trust factor. 

In human-human trust, in a situation in which trustee and trustor have similar purpose or intent could also evoke emotive/affective trust  \cite{johnson2005cognitive}, which is contingent on personal preference\cite{verberne2012trust}. Personal preference is shaped by culture and experiences \cite{schultz2002environmental} and culture is seen as a component that could affect trust in technology \cite{schaefer2016meta}. In human-computer trust, when a person finds a system agreeable or suits their personal taste, strong preference for the system and emotional attachment could be formed and affect the person trust \cite{madsen2000measuring}. 

The cross-discipline literature above identifies various concepts and factors that could affect trust in different contexts. However, even though multi-disciplinary source have added valuable insights to the construct, the diversity in wording and terms used should be unified. Therefore, we merged factors with similar meaning, and six trust factors were chosen to be the domains of the initial measurement instrument. The Table \ref{tab:trust} below summarises our definition of these domains, which are based on the literature. We merged domains that overlapped in meaning and modified some of their descriptions into the final six trust metrics: perceived understandability, perceived reliability, faith, perceived technical competence, perceived helpfulness, and personal attachment. 

Perceived Technical competence means that the system is perceived to perform the tasks accurately and correctly, based on the input information. Perceived Understandability means that user can form a mental model and predict future system behaviours. Perceived Reliability means that the system is perceived to be functioning consistently. Perceived helpfulness means that the system is perceived to provide adequate, effective, and responsive help. Faith means that the user is confident in the future ability of the system to perform, even in situations in which has never used the system before. Finally, personal attachment means that users find using the system agreeable, and consistent with their personal taste. 

\begin{table}
  \caption{Human-AI Trust Measurement Domains: Trust Factors and Descriptions}
  \label{tab:trust}
  \begin{tabular}{|l|p{8cm}|}
    \hline
    Trust Factors & Description\\
    \hline
    perceived technical competence  &  system is perceived to perform the tasks accurately and correctly based on the information that is input.\\\hline
    perceived reliability  &  system is perceived to be, in the usual sense of repeated, consistent functioning.\\\hline
    perceived understandability  &  user can form a mental model and predict future system behaviour. \\\hline
    personal attachment & user finds using the system agreeable, preferable, suits their personal taste. \\\hline
    faith & user has faith in the future ability of the system to perform even in situations in which it is untried. \\\hline
    perceived helpfulness  &  system is perceived to provide adequate, effective, and responsive help. \\\hline
  \end{tabular}
\end{table}

\subsubsection{Inductive Method\\}
\noindent\textbf{\underline{Online Survey}}\\
As mentioned previously, an online survey was conducted to help explore other possible trust factors that were not considered previously.  The online survey also aimed to test if any domain we included is not considered relevant or important by the general public in a healthcare context. We developed a dramatising vignette as a tool to help contextualise the measurement domains and items. The vignette technique is a method that can elicit perceptions, opinions, beliefs and attitudes from responses or comments to stories depicting scenarios and situations \cite{barter1999use}. Vignette is also appropriate to elicit users’ feedback on the implications of possible futures yet to be realised. Because AI assisted healthcare for preliminary cancer diagnosis that is able to provide explanation does not yet exist, this method is suitable for our study. 

\begin{table}[h]
\centering
\begin{tabular}{|l|l|l|l|}
\hline
\textbf{Variable} & \multicolumn{1}{c|}{\textbf{Value}} & \textbf{Frequency} & \textbf{\%} \\ \hline
\multirow{2}{*}{Gender} & Male & 25 & 47.1 \\ \cline{2-4} 
 & Female & 28 & 52.8 \\ \hline
\multirow{3}{*}{\begin{tabular}[c]{@{}l@{}}Age\\ (median = 35)\end{tabular}} & \textless{}30 & 16 & 30.1 \\ \cline{2-4} 
 & 30-40 & 23 & 43.3 \\ \cline{2-4} 
 & \textgreater{}40 & 14 & 26.4 \\ \hline
Total &  & 53 &  \\ \hline
\end{tabular}
\caption{Initial Survey Demographic}
\label{table:initial}
\end{table}


The demographic of initial online survey participants can be seen in Table \ref{table:initial}. We asked participants to first read a dramatising vignette and then to reflect on it and write down the potential main issues/problems (challenges) and the potential main advantages/benefits (opportunities) of AI systems in healthcare. The potential issues pointed out by participants mostly fell under \textit{Competency} group, such as, misdiagnosis, false positive, false negative, or inaccuracy (34 instances). This finding can be put as a supporting argument that perceived technical competence is an important consideration for user to use an AI system. 
In the relation to explanation, participants mentioned their concern on the lack of information or information that is not laypeople friendly. \textit{"Yes. More information is always better for a diagnosis"- P24. "Vague information or terms an average person may not use,..."- P52.} This finding can be linked to the perceived understandability and perceived helpfulness domains. 

For the potential main advantages/benefits (opportunities) of AI system in healthcare, specifically AI breast cancer self-diagnosis system, participants mentioned \textit{Fast Result} (18 instances) compared to the traditional diagnosis process. \textit{"Get some answers quickly, which gives you peace of mind"- P20.} The AI system could also offers information which can be beneficial for users as a \textit{Second Opinion} (10 instances) provider, or provider for \textit{More Information} (9 instances) in general, or both. \textit{"If used regularly, it can also be a great tool to track changes in your condition without having to see a doctor."- P50.} 

The topic of trust came up several times (13 instances), specially disproportionate level of trust, such as, \textit{Over-trusting} (7 instances) and \textit{Distrust} (6 instances) on the AI system. \textit{"People could put too much weight on the result giving it the same value as a true diagnosis."-P12. "People don't trust it enough."-P48.} However, overall, we found no new dimension of trust factor from the open-ended answers. Since the nature of free-text survey response is often lack attention to context and conceptual richness \cite{o2004any}, the answers rarely produce a rich qualitative data. 

We also did measurement instrument pre-testing, as a part of domains and items selection and/or reduction, and also to test the form of measurement instrument. We created the initial measurement instrument based on the literature with six domains of trust factor and three statement items for each domain. Based on our knowledge, there is no rule of thumb for the number of items should be included in the measurement scale, as long as it's not too long that inspired participation fatigue or motivation \cite{schultz2005measurement}. Since this initial online survey is quite long, we decided to only use three statements for each domain, making it 18 statement in total. Participants were asked to rate, in Likert 7-point scale, the importance of 18 item statements before and after reading the dramatising vignette. For example, to evaluate perceived technical competence, participants were asked to rate the following statement: "The application would use appropriate methods to get results based on the information I input."(See Appendix A). 

We inspected mode, median, and mean values for each item and internal consistency of each domain was then measured, using Cronbach’s alpha (See Table \ref{table:rating}). Based on the median rating, most of the domains are rated as very (rating = 6) or extremely (rating = 7) important by the survey respondents. The only item rated negatively (rating < 4) was from personal attachment domain, with the statement: "\textit{you feel a sense of loss if the app is suddenly unavailable to use}". Perceived reliability, perceived technical competence, and perceived helpfulness domains demonstrated excellent internal consistency, with their alpha coefficient > 0.8 \cite{cronbach1951coefficient}\cite{nunnally1994psychometric}. Meanwhile, perceived understandability, personal attachment, and faith domains' internal consistency can be regarded as acceptable. However, the overall initial measurement is still demonstrated excellent reliability Cronbach’s alpha > 0.94. From this result, we argued that the initial measurement scale is a good starting point, with some refinement need to be done in perceived understandability, personal attachment, and faith items. 

\begin{table*}[]
\makebox[\textwidth]{
\begin{tabular}{|l|lll|lll|lll|lll|lll|lll|}
\hline
\multirow{2}{*}{} & \multicolumn{3}{c|}{reliability} & \multicolumn{3}{c|}{\begin{tabular}[c]{@{}c@{}}technical \\ competence\end{tabular}} & \multicolumn{3}{c|}{understandability} & \multicolumn{3}{c|}{\begin{tabular}[c]{@{}c@{}}personal \\ attachment\end{tabular}} & \multicolumn{3}{c|}{helpfulness} & \multicolumn{3}{c|}{faith} \\ \cline{2-19} 
 & \multicolumn{1}{l|}{r1} & \multicolumn{1}{l|}{r2} & r3 & \multicolumn{1}{l|}{tc1} & \multicolumn{1}{l|}{tc2} & tc3 & \multicolumn{1}{l|}{u1} & \multicolumn{1}{l|}{u2} & u3 & \multicolumn{1}{l|}{p1} & \multicolumn{1}{l|}{p2} & p3 & \multicolumn{1}{l|}{h1} & \multicolumn{1}{l|}{h2} & h3 & \multicolumn{1}{l|}{f1} & \multicolumn{1}{l|}{f2} & f3 \\ \hline
mode & \multicolumn{1}{l|}{7} & \multicolumn{1}{l|}{7} & 7 & \multicolumn{1}{l|}{7} & \multicolumn{1}{l|}{7} & 7 & \multicolumn{1}{l|}{7} & \multicolumn{1}{l|}{7} & 7 & \multicolumn{1}{l|}{2} & \multicolumn{1}{l|}{5} & 5 & \multicolumn{1}{l|}{7} & \multicolumn{1}{l|}{7} & 6 & \multicolumn{1}{l|}{5} & \multicolumn{1}{l|}{7} & 6 \\ \hline
median & \multicolumn{1}{l|}{7} & \multicolumn{1}{l|}{7} & 7 & \multicolumn{1}{l|}{7} & \multicolumn{1}{l|}{7} & 7 & \multicolumn{1}{l|}{6} & \multicolumn{1}{l|}{7} & 6 & \multicolumn{1}{l|}{3} & \multicolumn{1}{l|}{5} & 5 & \multicolumn{1}{l|}{7} & \multicolumn{1}{l|}{6} & 6 & \multicolumn{1}{l|}{5} & \multicolumn{1}{l|}{6} & 6 \\ \hline
mean & \multicolumn{1}{l|}{6.2} & \multicolumn{1}{l|}{6.0} & 6.4 & \multicolumn{1}{l|}{6.3} & \multicolumn{1}{l|}{6.2} & 6.0 & \multicolumn{1}{l|}{5.6} & \multicolumn{1}{l|}{6.0} & 5.7 & \multicolumn{1}{l|}{3.0} & \multicolumn{1}{l|}{5.2} & 5.3 & \multicolumn{1}{l|}{6.1} & \multicolumn{1}{l|}{5.8} & 5.7 & \multicolumn{1}{l|}{5.2} & \multicolumn{1}{l|}{5.5} & 5.4 \\ \hline
 \(\alpha\) & \multicolumn{3}{l|}{0.81} & \multicolumn{3}{l|}{0.82} & \multicolumn{3}{l|}{0.77} & \multicolumn{3}{l|}{0.75} & \multicolumn{3}{l|}{0.90} & \multicolumn{3}{l|}{0.76} \\ \hline
\end{tabular}}
\caption{Importance Rating: Initial Measurement Scale with Six Domains and Three Statement Items Each}
\label{table:rating}
\end{table*}

\noindent \textbf{\underline{Focus Group}}\\
The focus group analysis was used to capture any missing human-AI trust factor that was not captured by the literature. During the focus group participants were asked to read the dramatising vignette and then have an open discussion on human-AI trust and the factors affecting this relationship.The codes emerged can be grouped to these core variables: User Needs, Communication, User Concern, AI usage, and Trust. In the following, we describe one of the core variable: Trust, as a part of inductive method in this item generation phase. 

Based on the questions asked about trust before and after the vignette, trust is affected by communication and credibility. AI systems' credibility could be proven with license or certification. License and certification are required for all medical tools, and AI in healthcare should too.\textit{"Overseeing bodies, both in the U.S. and here, and elsewhere. I think (here) it’s BMC, the royal colleges. Things that you have to be able to practice medicine in most countries. Integrating AI into that system somehow. Whether it's, through having to release your bug report and knowing what the control condition is you run the amazing test results. "-P5}.\textit{"What if AI goes through a residency with real physicians. The tool itself needs to be continuously improved for a period of two years by physicians, by experts, in clinical practice."-P3}\\

Autonomy is an important principle in medical ethics as perceived by patients, and it means individuals demand to be free to choose whether and what kind of treatment to receive \cite{jonsen1982clinical}. Autonomy is relevant for both interactions between medical professionals and patients, and between AI systems and users. In the AI healthcare system context, users should have the right to make decisions for themselves; and should be put in the right conditions that enable them to make those decisions in a well informed but autonomous way. The decisions mentioned by participants vary from decisions regarding treatment, to the decision regarding whether or not they want to use the system, or even decisions about the conditions that enable them to make decisions, in this case, is the decision about what kind of information users want to be given in the explanation. \textit{"I would like to be able to invoke it or turn it off at my choice."-P1."you could be given references if you wanted to do research. but it's also up to you"-P6}. \\

From these results, two main additional trust factors emerged: \textit{Institutional Credibility} and \textit{User Autonomy} (See Table \ref{tab:trust2}). For \textit{Institutional Credibility}, participants meant the users’ trust is placed in the institution which regulates the certification of the AI system. We previously did not include institution-based trust in the initial domain, because according to McKnight et al.’s paper, institution-based trust is outside of the overall concept of trust in specific technology \cite{mcknight2011trust}. Since trust towards an AI system can be considered as trust in specific technology, we did not include institutional credibility, or structural assurance supporting the technology use. 

\textit{User Autonomy} is also deemed as important by participants. As mentioned above, user should be able to control their decision regarding treatment or regarding whether or not to use the system. This is in line with previous research on trust in healthcare, patient's trust will improve when doctors give patient autonomy by letting them manage their disease \cite{rowe2006trust} \cite{croker2013factors}. 

\begin{table}[h]
  \caption{Human-AI Trust Measurement Additional Domains: Trust Factors and Descriptions}
  \label{tab:trust2}
  \begin{tabular}{|l|p{10cm}|}
    \hline
    Trust Factor & Description  \\
    \hline
    institutional credibility  &  user beliefs that the technology has been tested or certified by overseeing bodies. \\\hline
  user autonomy & user knows and able to decide for their own decision.  \\\hline
  \end{tabular}
\end{table}

\section{Item Evaluation}
\subsection{Methods}
After the initial development process, we further review and revise initial domains and items.  Since it is recommended to create a large number of items in the early stage of item development \cite{hinkin1998brief}, we created five item statements for each domain, including two new added domains, as our second version of measurement instrument (See Appendix). To assess the new measurement instrument, an evaluation by experts and target population was carried out. 

\subsubsection{Expert Validation}
Evaluation by expert entails analysis of content validity, which is the degree to which items of a measurement instrument are relevant to and representative of the targeted construct for a particular measurement purpose \cite{cook2006current,devellis2021scale}. This is one of the most important of measurement instrument validations \cite{boateng2018best}. Content validity is also used to observe the correct grammar and appropriate wording in items and appropriate scoring \cite{safikhani2013qualitative}. In summary, experts review the clarity (the question is clear and specific. the question make sense to the reader), coherence (the question is logical, consistent, and reasonable in the context of the research problem being addressed), and completeness (the statement presented is fully represented by the definition) of the measurement instrument items \cite{quiroz2017development}. To conduct expert validation, we went through several stages: content validation form development, expert selection, validation administration, and expert rating analysis. 

Similar to any measurement instrument, the content validation form should be developed with the appropriate items to allow the experts to have a clear expectation and understanding about the task. As defined above, content validity is the degree of relevance and representativeness \cite{cook2006current} and clarity of the targeted domain item \cite{safikhani2013qualitative}. Therefore, the form should include at least two evaluations: content relevancy and content clarity. We presented the domains and its definitions, followed by five items related to them to the experts. The experts were requested to critically review the domain and its items before providing rating on each item using a 5-point Likert scale of agreement. For each item, we asked if "\textit{the statement is clear, consistent, specific, and a non-expert reader would be able to make sense of it}" to provide evidence of content clarity, and if \textit{"the statement fully represents the definition of perceived understandability}" to provide evidence of content representativeness and relevancy. Additionally, we asked experts if they understand the domain definition because unambiguously defined domain can also be considered as one of conditions of content validity \cite{guion1977content}. Since the measurement instrument review requires verbal comments from experts, a mixed data collection method was administered, where qualitative data was collected via interview and quantitative data was collected via content validation form. 

The next stage is experts selection. The experts involved in this stage should be highly knowledgeable about the domain of interest and measurement instrument development \cite{devellis2021scale}\cite{morgado2017scale}. Since this measurement instrument was developed in the context of AI systems in healthcare, the domain includes; AI/computing and medicine/healthcare. Therefore, we aimed to recruited AI/computing experts and medical experts, in addition to measurement instrument development expert. 

Invitations for interview were sent via email to possible experts via email from our personal research network. The number of expert recommended by the literature is two at the minimum \cite{davis1992instrument} and ideally range between five to seven experts \cite{haynes1995content}\cite{polit2006content}. In the end, we selected seven experts: two scale development experts with psychology background, two scale development experts from computing field, one AI expert, and two medical experts. The interviews ran for one hour and were transcribed in the process. 

The last stage of expert validation is expert rating analysis. The Content validity index (CVI) was calculated to measure proportional agreement \cite{lynn1986determination}\cite{polit2006content}.  Although CVI is broadly used for content validity, this index has been criticised for not considering possible inflated values caused by chance agreement \cite{wynd2003two}. Therefore, we also calculated Cohen's coefficient kappa (\(\kappa\)) \cite{cohen1960coefficient} which adjusts for chance agreement and is considered as the most efficient \cite{wynd2003two}. Items with low value of CVI and \(\kappa\) were considered invalid and removed from the measurement instrument.  Lastly, comments from experts worked as a guide and consideration to refine the trust factors and revise the rest of the items.

\subsubsection{Cognitive Interview}
Not only from experts, evaluation on the target population is also important. We conducted Cognitive Interview, to evaluate if the items reflect the domain of study by ensuring that the target population understand the item statements and/or questions\cite{beatty2007research} and help refine the measurement instrument items. Cognitive Interview can help improve clarity, identify confusing and problematic item, indicate problematic item order, reveal the thought process of participants, ensure the intended data are produced \cite{willis2004cognitive}\cite{tourangeau2003cognitive}, and is the recommended method to evaluate the measurement instrument before Survey Administration \cite{boateng2018best}. 

It is recommended to run 5-15 interviews from target population sample \cite{willis2004cognitive}\cite{beatty2007research}. The interview technique combined think-aloud approach with verbal probing. In the interview, first, we described the study and introduced participants to AI technology and AI in healthcare. Since, the target population is non-experts/laypeople this introduction process is important. Participants were then asked to read the item statement, answer if they can understand and make sense of it, and explain what does the statement mean using their own words. Based on the answer, verbal probing might occur.  Invitations for this interview were sent via email and electronic messages from our personal and professional network. 

\subsection{Results}
\subsubsection{Expert Validation}
We conducted interviews with seven experts separately. The experts include professionals and academics: two scale development experts from psychology field, two scale development experts from computing field, one AI expert, and two medical experts. In the first round, CVI was calculated in item level by dividing the number of experts giving a rating 4 or 5 to the representativeness of each item with the total number of experts. Evaluation criteria for CVI is "Excellent" for CVI \(>\) 0.79 \cite{davis1992instrument,seif2004educational,polit2006content} and "For Revision" for CVI \(>\) 0.7 \cite{abdollahpour2010process}. After CVI for all instrument items were calculated, kappa was calculated using numerical values of probability of chance agreement (PC) and CVI of each item in following formula:

\[K= (CVI - PC) / (1- PC).\]
Evaluation criteria for kappa value are "Excellent" for \(\kappa \geq \)  0.74, "Good" for 0.74\(> \kappa \geq\)0.6, and "Fair" for 0.59\(> \kappa \geq\)0.40 \cite{cicchetti1981developing}. Among the 40 instrument items, 16 items with CVI lower than 0.7 and \(\kappa\) score lower than 0.4 were interpreted as "Invalid" and removed from the measurement instrument (See Table \ref{table:ev1} and Table \ref{table:ev2} ). However, the number of items removed from each dimension were not equal and range from three items (\textit{perceived understandability} and \textit{perceived helpfulness}) to one item (\textit{user autonomy} and \textit{faith}). Thus, to have the same number of item for each domain, we selected the two best items from each domain. When domain has two or more items passed the CVI and \(\kappa\) "Excellent" threshold, items with "For Revision" scores were removed, making it 20 items in the end. 

In the second round, we looked at the Cohen's \(\kappa\) of the Clarity ratings from the rest of 20 items, to decide if the item needs revision or not. The evaluation criteria for kappa value is the same as above. Items with \(\kappa <\)0.4 were considered "Poor" and removed from the item pool, and items with "Excellent" clarity were accepted without major revision. Since only two best items were selected for each domain, one of \textit{faith} items with "Fair" clarity and one of \textit{user autonomy} items with EA \(<\) 7 were removed. In the end, we have the measurement instrument consists of 16 items from 8 domains, with comments from experts that helped refine and revise the items accordingly.

\begin{table}[]
\makebox[\textwidth]{
\begin{tabular}{|llllllll|}
\hline
\multicolumn{1}{|l|}{Dimension Items} & \multicolumn{1}{c|}{EA} & \multicolumn{1}{c|}{CVI} & \multicolumn{1}{c|}{PC} & \multicolumn{1}{c|}{\(\kappa\)} & \multicolumn{1}{c|}{\begin{tabular}[c]{@{}c@{}}Interpretation\\ (CVI)\end{tabular}} & \multicolumn{1}{c|}{\begin{tabular}[c]{@{}c@{}}Interpretation\\ (\(\kappa\))\end{tabular}} & \multicolumn{1}{c|}{Clarity} \\ \hline
understandability &  &  &  &  &  &  &  \\
u1 & 7 & 1.000 & 0.008 & 1.000 & Excellent & Excellent & Excellent \\
u2 & 6 & 0.857 & 0.055 & 0.849 & Excellent & Excellent & Excellent \\
u3 & 1 & 0.143 & 0.055 & 0.093 & Invalid & Invalid &  \\
u4 & 4 & 0.571 & 0.273 & 0.410 & Invalid & Fair &  \\
u5 & 3 & 0.429 & 0.273 & 0.214 & Invalid & Invalid &  \\ \hline
technical competence &  &  &  &  &  &  &  \\
tc1 & 6 & 0.857 & 0.055 & 0.849 & Excellent & Excellent & Excellent \\
tc2 & 5 & 0.714 & 0.164 & 0.658 & For Revision & Good & Poor \\
tc3 & 3 & 0.429 & 0.273 & 0.214 & Invalid & Invalid &  \\
tc4 & 5 & 0.714 & 0.164 & 0.658 & For Revision & Good & Excellent \\
tc5 & 4 & 0.571 & 0.273 & 0.410 & Invalid & Fair &  \\ \hline
reliability &  &  &  &  &  &  &  \\
r1 & 6 & 0.857 & 0.055 & 0.849 & Excellent & Excellent & Fair \\
r2 & 5 & 0.714 & 0.164 & 0.658 & For Revision & Good & Excellent \\
r3 & 5 & 0.714 & 0.164 & 0.658 & For Revision & Good & Poor \\
r4 & 2 & 0.286 & 0.164 & 0.146 & Invalid & Invalid &  \\
r5 & 4 & 0.571 & 0.273 & 0.410 & Invalid & Fair &  \\ \hline
helpfulness &  &  &  &  &  &  &  \\
h1 & 7 & 1.000 & 0.008 & 1.000 & Excellent & Excellent & Excellent \\
h2 & 6 & 0.857 & 0.055 & 0.849 & Excellent & Excellent & Good \\
h3 & 3 & 0.429 & 0.273 & 0.214 & Invalid & Invalid &  \\
h4 & 1 & 0.143 & 0.055 & 0.093 & Invalid & Invalid &  \\
h5 & 4 & 0.571 & 0.273 & 0.410 & Invalid & Fair &  \\ \hline
\end{tabular}}
\caption{Expert Validation Review Ranking: Dimension Items' Experts in Agreement (EA), Content Validity Index (CVI), Probability of Chance agreement (PC), Cohen's coefficient kappa \(\kappa\), CVI evaluation interpretation, \(\kappa\) evaluation interpretation, \(\kappa\) item Clarity (Part 1)}
\label{table:ev1}
\end{table}

\begin{table}[]
\makebox[\textwidth]{
\begin{tabular}{|llllllll|}
\hline
\multicolumn{1}{|l|}{Dimension Items} & \multicolumn{1}{c|}{EA} & \multicolumn{1}{c|}{CVI} & \multicolumn{1}{c|}{PC} & \multicolumn{1}{c|}{\(\kappa\)} & \multicolumn{1}{c|}{\begin{tabular}[c]{@{}c@{}}Interpretation\\ (CVI)\end{tabular}} & \multicolumn{1}{c|}{\begin{tabular}[c]{@{}c@{}}Interpretation\\ (\(\kappa\))\end{tabular}} & \multicolumn{1}{c|}{Clarity} \\ \hline
personal attachment &  &  &  &  &  &  &  \\
pa1 & 7 & 1.000 & 0.008 & 1.000 & Excellent & Excellent & Excellent \\
pa2 & 6 & 0.857 & 0.055 & 0.849 & Excellent & Excellent & Good \\
pa3 & 5 & 0.714 & 0.164 & 0.658 & For Revision & Good &  \\
pa4 & 4 & 0.571 & 0.273 & 0.410 & Invalid & Fair &  \\
pa5 & 3 & 0.429 & 0.273 & 0.214 & Invalid & Invalid &  \\ \hline
user autonomy &  &  &  &  &  &  &  \\
ua1 & 7 & 1.000 & 0.008 & 1.000 & Excellent & Excellent & Excellent \\
ua2 & 6 & 0.857 & 0.055 & 0.849 & Excellent & Excellent & Excellent \\
ua3 & 7 & 1.000 & 0.008 & 1.000 & Excellent & Excellent & Excellent \\
ua4 & 5 & 0.714 & 0.164 & 0.658 & For Revision & Good &  \\
ua5 & 4 & 0.571 & 0.273 & 0.410 & Invalid & Fair &  \\ \hline
faith &  &  &  &  &  &  &  \\
f1 & 6 & 0.857 & 0.055 & 0.849 & Excellent & Excellent & Fair \\
f2 & 6 & 0.857 & 0.055 & 0.849 & Excellent & Excellent & Excellent \\
f3 & 7 & 1.000 & 0.008 & 1.000 & Excellent & Excellent & Excellent \\
f4 & 4 & 0.571 & 0.273 & 0.410 & Invalid & Fair &  \\
f5 & 5 & 0.714 & 0.164 & 0.658 & For Revision & Good &  \\ \hline
institution credibility &  &  &  &  &  &  &  \\
ic1 & 6 & 0.857 & 0.055 & 0.849 & Excellent & Excellent & Good \\
ic2 & 6 & 0.857 & 0.055 & 0.849 & Excellent & Excellent & Good \\
ic3 & 1 & 0.143 & 0.055 & 0.093 & Invalid & Invalid &  \\
ic4 & 4 & 0.571 & 0.273 & 0.410 & Invalid & Fair &  \\
ic5 & 5 & 0.714 & 0.164 & 0.658 & For Revision & Good &  \\ \hline
\end{tabular}}
\caption{Expert Validation Review Ranking: Dimension Items' Experts in Agreement (EA), Content Validity Index (CVI), Probability of Chance agreement (PC), Cohen's coefficient kappa \(\kappa\), CVI evaluation interpretation, \(\kappa\) evaluation interpretation, \(\kappa\) item Clarity (Part 2)}
\label{table:ev2}
\end{table}

\subsubsection{Cognitive Interview}
We conducted qualitative interviews with nine participants. All participants are laypeople, with age range: six participants were below 30, three participants were in 30-45, and three participants were above 45 years old. Each cognitive interview lasted one to two hours in semi-structured format. As described above, participants' were expected to think-aloud their understanding on each item statement using their own words. Since no specific data analysis method is recommended by cognitive interview literature \cite{willis2004cognitive}\cite{beatty2007research}, general view on participants' understanding and in-depth look on participants' cognitive processing were noted. 

In general, participants claimed that the items are understandable and make sense. When participants explained their interpretation on the item statements, the description of their mental process allowed participants to answer in a manner that reflects their experience, which indicates their understanding \cite{willis2004cognitive}. The participants were able to understand correctly the specifications of the items and, crucially, the interpretation were consistent across participants reflect the trust factor definition. 

However, some inconsistencies between participants' interpretation were found on two item statements from perceived technical competence and personal attachment domains. One statement from perceived technical competence: "\textit{The AI system uses appropriate methods to get results and to reach decisions based on the information I input}" was understood differently. The cause of this inconsistency was the word "appropriate". Some interpretations on "appropriate method" were: ethical method, method that gets the job done, method human professionals uses, or method where the data is taken without the user's permission. One of the participants claimed that the word appropriate is "a bit too fluid". 

In the statement from the personal attachment factor,  \textit{"I find the AI system suitable for my style and I would feel a sense of loss if I could no longer use it."}, the word "style" was interpreted differently. As described previously, personal attachment measures the degree to which the user finds using the system agreeable, preferable, and suits their personal taste. The sentence was developed based on the style's definition from Cambridge Dictionary as a way of doing something, especially one that is typical of a person. Even though some participants interpreted style correctly, most participants interpreted style as lifestyle. Thus, when they reflected on their experience, they drove it from their lifestyle. " I think style is lifestyle, the app is suitable for my lifestyle. Like I like to do yoga so the apps that suit my style are health yoga app. "- P8. One of the participants even appeared confuse when reading the statement. "Hmm. That's an interesting one. Yeah. Suitable for my style. And I would feel a sense of loss if I could no longer use it. Um, “my style” is intriguing the way you worded that."-P2. 

Lastly, the word "vendor" in the institution credibility statement was found to be confusing.  \textit{"I am confident in the AI system capability because it is developed by a reputable institution, and backed by valid vendors and consumer protections."}, "I'm not sure what vendor means. [...] Vendors makes me think of food."-P5. Based on this result, none of the item statements were dropped and some modifications on the word choice were applied. 

\section{Survey Administration}
\subsection{Method}
After all item statements and domains were evaluated, we administered the main survey to further evaluate the measurement instrument. A survey could identify the characteristics of a broad population of individuals if a clear research question inquiring about the nature of the target population is present \cite{easterbrook2008selecting}. We chose to use an online survey, because an online survey takes advantage of the Internet to provide access to broader groups and efficient in time \cite{wright2005researching}. 

The online survey consists of; demographic questions (gender and age group), their general trust towards AI in healthcare, their likelihood to use AI in healthcare, and the trust measurement instrument at the end (See Table \ref{table:survey}). Participants were asked to read the information page and filled the consent form before proceeding to the online survey. The measurement instrument was presented after videos of available AI medical support systems. The online survey contains two sets of questions with 20 items in total. The first set of questions contains two demographic items (gender and age group) and two trust propensity questions. The second set of questions contains 16 statements from the measurement instrument. Between the first and the second set of questions, participants were assigned to watch a two-minutes video of cancer detection/risk assessment applications available on the market, such as, SkinVision (skin cancer detection), Braster (breast cancer detection), and Alexa Babylon (health assessment). Participants were then asked to rate their agreement with the statements from the measurement instrument based on the application that they just saw using 7-point Likert scales. The online survey was developed using the Google Forms platform and published via Amazon Mechanical Turk. To minimise submission from bots, we only accept master worker and put different codes in the survey to submit at the end.

To decide the sample size or the number of participant, we looked at the guidelines given by the literature. The overall recommendation is the larger sample size/participants the better. The number of participants can be determine using the ratio of participant and instrument item, such as, 5:1 participant-to-item ratio \cite{gorsuch1988exploratory} and 10:1 participant-to-item ratio \cite{nunnally1967psychometric}. The number of participants can also be determined without taking the number of item into account. The 200-300 range is argued to be appropriate \cite{guadagnoli1988relation}, and other literature considered 300 participants as good \cite{comrey1988factor}\cite{clark1995constructing}. Thus, we aimed to have 300 participants. 

\begin{table}[]
\makebox[\textwidth]{
\begin{tabular}{l|p{12cm}}
\hline
Domain & Statement \\\hline
Understandability & I understand how the AI system works and I feel confident I will be able to use it in the future. \\
 & I understand how the AI system behaves, how it can assist me, and what I can expect from using it in the future. \\\hline
Technical Competence & The AI system uses appropriate methods to get results based on the information I input. \\
 & The AI system correctly uses the information I input to provide accurate results. \\\hline
Reliability & The AI system consistently provides the results it is expected to produce. \\
 & The AI system responds the same way under the same conditions at different times. \\\hline
Helpfulness & When I need help, the AI system responds to my needs effectively and responsively. \\
 & The AI system provides me with the effective and responsive help I need. \\\hline
Personal Attachment & I find the AI system suits my preference and I would feel a sense of loss if I could no longer use it. \\
 & I like using the AI system because it suits me, and always want to use it. \\\hline
User Autonomy & I feel in control when operating the various functions and features of the AI system. \\
 & The AI system has functionalities and features I can control. \\\hline
Faith & When I am unsure about the AI system’s result, I believe in the AI system rather than myself. \\
 & Even if I am not sure about the result and the actual performance, I am confident that the AI system will provide the best result. \\\hline
Institutional Credibility & I feel assured using the AI system because it is made by a reputable institution and therefore already went through a credible regulation process. \\
 & I am confident in the AI system capability because it is developed by a reputable institution, and backed by valid companies and consumer protections. \\\hline
\end{tabular}}
\caption{16 Survey Questions/Statements (Two Statements for Each Domain}
\label{table:survey}
\end{table}

\subsection{Results}
The Mechanical Turk tasks were up for two weeks and data from 300 participants were collected. The participants were 52.7\% male, 47\% female, 0.3\% prefer not to say their gender. Almost half of the participants were between 30-40 years old (46\%), with the rest 22\%  were between 40-50 years old, 18.7\% between 20-30 years old, 13\% above 50 years old, and 0.3\% below 20 years old. Further analyses for measurement instrument evaluation, methods and results, are described in the following sections. To determines if the responses given with the sample are adequate, we performed the Kaiser–Meyer–Olkin (KMO) test \cite{kaiser1958varimax}. KMO measures the sampling adequacy (MSA) with criterion: 0-0.49 as unacceptable, 0.50-0.59 as miserable, 0.60-0.69 as mediocre., 0.70-0.79 as middling, 0.80-0.89 as meritorious, and 0.90-1.00 as marvelous \cite{kaiser1958varimax}. Additionally, we ran Bartlett's test of sphericity test \cite{bartlett1954note} to test the null hypothesis: the correlation matrix is an identity matrix. If the significance level is less than 0.05, the null hypothesis is rejected, which means the items are suitable for structure detection. As depicted in Table \ref{table:kmo}, KMO results of high value (0.9458) implied the adequacy of the sampling data and a significant test statistic (0.00) by Bartlett’s test of sphericity indicated that a factor analysis may be useful with this data. 

\begin{table}[h]
\begin{tabular}{lcc}
\hline
Kaiser-Meyer-Olkin Measure of Sampling Adequacy & \multicolumn{1}{l}{} & 0.9458 \\ \hline
\multirow{3}{*}{Bartlett’s Test of Sphericity} & Approx. Chi-Square & 3634.78 \\
 & df & 120 \\
 & Sig. & 0.00 \\ \hline
\end{tabular}
\caption{Kaiser–Meyer–Olkin (KMO) sample adequacy test.}
\label{table:kmo}
\end{table}

\section*{Measurement Instrument Evaluation}
\section{Test of Dimensionality}
\subsection{Method}
To test the dimension of the proposed measurement instrument, we used Factor Analysis. Two main classes of factor analysis are Exploratory Factor Analysis (EFA) and Confirmatory Factor Analysis (CFA). EFA seeks to discover the measurement model and, as the name suggest, exploratory in nature. Meanwhile, CFA starts with a hypothesis model, such as, how many factors and which items load on which factors. Our current model consists of eight (8) dimensions of trust factors: perceived reliability, perceived technical competence, perceived understandability, faith, personal attachment, perceived helpfulness, user autonomy, and institution credibility; with two items on each dimension. Given that we have hypothesised the trust model, confirmatory factor analysis (CFA) was performed in this stage. CFA investigates how well the hypothesised factor structure (model) fits with the data \cite{mackenzie1991organizational}. Some of the fit indices are: Comparative Fit Index (CFI \textgreater 0.95), Tucker Lewis Index (TLI \textgreater 0.95), Root Mean Square Error of Approximation (RMSEA \textless 0.06), Standardized Root Mean Square Residual (SRMR \textless 0.08) and low Chi-square \cite{hu1999cutoff} \cite{hinkin1998brief}\cite{boateng2018best}.

\subsection{Results}
Table \ref{table:cfa} shows the fit indices for the hypothesised model. Based on the root mean square error of approximation (RMSEA), standardized root mean square residual (SRMR), comparative fit index (CFI), and Tucker–Lewis index (TLI) reported, the hypothesised model fits well and does not need additional alteration. 

\begin{table}[h]
\begin{tabular}{ll}
\hline\\\hline
Chi-square & 124.209\\
Comparative Fit Index (CFI) & 0.987 \\
Tucker-Lewis Index (TLI) & 0.979 \\
Root Mean Square Error of Approximation (RMSEA) & 0.046 \\
Standardized Root Mean Square Residual (SRMR) & 0.023 \\ \hline
\end{tabular}
\caption{Confirmatory Factor Analysis of Trust Model}
\label{table:cfa}
\end{table}


\section{Test of Reliability}
\subsection{Methods}
\subsubsection{Internal Consistency}
As mentioned previously, reliability refers to the degree of consistency demonstrated when a measurement is repeated under the same conditions \cite{porta2014dictionary}. Multiple assessments of reliability have been developed, and among these assessment methods, the coefficient alpha, specifically Cronbach's alpha \cite{cronbach1951coefficient}, has been the most widely used measure of reliability. Coefficient alpha is commonly used to estimate one type of reliability: internal consistency. Internal consistency represents the degree to which the measurement items are inter-correlated, or if they assess the same construct consistently. However, many studies have criticised coefficient alpha, pointing out that coefficient alpha can only be treated as scale reliability when specific conditions hold \cite{raykov2015direct},which unlikely to hold in practice\cite{cho2015cronbach,green2009commentary}. These conditions are: (1) items are unidimensional; (2) the average factor loading is above .7;  (3) the differences between individual factor loadings and average factor loading are less than .2 \cite{raykov2015direct}. Failure to hold these three conditions could resulted in biased result. 

The proposed alternative to coefficient alpha is coefficient omega, which has been argued to be a more sensible measurement of internal consistency \cite{raykov2015direct,green2009commentary}. Moreover, the formulation of coefficient omega is deemed to match the definition of reliability \cite{mcdonald2013test}. However, the difference between coefficient omega and coefficient alpha was found to be small in applications \cite{maydeu2007asymptotically,raykov1997scale}, and coefficient alpha can be treated as identical to coefficient omega when the conditions (1-3) above hold \cite{raykov2015direct}.
Thus, we used both coefficient alpha and omega to assess the reliability of our measurement instrument, with Cronbach's alpha \cite{cronbach1951coefficient}, Raykov's \cite{raykov2011introduction}, Bentler's \cite{bentler2009alpha}, and McDonald's omega \cite{mcdonald2013test}. If the value of McDonald's omega is similar to the other two omegas (Bentler's and Raykov's), it indicates that the model fits the data well \cite{bentler2009alpha}.

\subsubsection{Test-retest reliability}
To assess the instrument temporal stability, test-retest reliability was conducted. Test-retest reliability looks at the reliability across time, in which the same participants are able to perform similarly in different times \cite{raykov2015direct}. The tests usually quantified using correlation \cite{cook2006current}, such as, Intra-class Correlation Coefficient \cite{streiner2015health} and Pearson product moment correlation coefficient (Pearson r) \cite{pearson1896vii}. Even though reliability was often assessed using Pearson r \cite{weir2005quantifying}, Pearson r was not recommended for assessing test-retest reliability \cite{kroll1962note}, especially for non-continuous data \cite{ludbrook2002statistical}. If the correlation value is high (close to 1), it indicates high test–retest reliability; and if the correlation value is close to zero, it indicates low reliability. 

\subsubsection{Replication: Repeated Survey}
The test-reliability is a part of replication processes, where the survey was administered in two (or more) different times to the same group of people. In order to do this, we collected additional data and repeated the online survey. The repeated survey consists of demographic questions (gender and age group), their general trust towards AI in healthcare, and the trust measurement instrument (See Table \ref{table:survey}). Participants were then asked to interact with AI healthcare prototypes and rate their agreement with the statements from the measurement instrument using 7-point Likert scale. The online survey was developed using the Google Forms platform and published via Amazon Mechanical Turk. Similar with the main survey, to minimise data from bots, we only accept master worker and put different codes in the survey to submit at the end.

\subsection{Results}
\subsubsection{Internal Consistency}
The coefficient omegas and coefficient alpha were calculated in R. Table \ref{table:reliable} depicts that all alphas and omegas values are above 0.7, indicating internal consistency in all dimensions and overall measurement\cite{nunnally1994psychometric}. Additionally, the results show that all Raykov's, Bentler's, and McDonald's coefficient omega are similar, suggesting that the model fits the data well. This support the finding in previous Test of Dimensionality stage.

\begin{table}[h]
\makebox[\textwidth]{
\begin{tabular}{lrrrr}
\hline
 & \multicolumn{1}{c}{Cronbach's alpha} & \multicolumn{1}{c}{Raykov's omega} & \multicolumn{1}{c}{Bentler's omega} & \multicolumn{1}{c}{Mcdonald's omega} \\\hline
reliability & 0.7232 & 0.7244 & 0.7244 & 0.7244 \\
technical competence & 0.8371 & 0.8383 & 0.8383 & 0.8383 \\
understandability & 0.7860 & 0.7938 & 0.7938 & 0.7938 \\
personal attachment & 0.8388 & 0.8393 & 0.8393 & 0.8393 \\
helpfulness & 0.8683 & 0.8699 & 0.8699 & 0.8699 \\
faith & 0.8285 & 0.8306 & 0.8306 & 0.8306 \\
user autonomy & 0.8289 & 0.8311 & 0.8311 & 0.8311 \\
institution credibility & 0.9136 & 0.9136 & 0.9136 & 0.9136\\
overall measurement & 0.9481 & 0.9512 & 0.9512 & 0.9504\\\hline
\end{tabular}}
\caption{Reliability tests for the measurement instrument.}
\label{table:reliable}
\end{table}

In relation to the different reliability assessment measures, the results shown that the difference in coefficient alphas and omegas are small (less than .1).  To see if this is the case where coefficient alpha can be treated as coefficient omega, we tested the three conditions of coefficient alpha. However, we could not conclude if the conditions hold. We fit the data to unidimensional model (single-factor model) with CFA and applying the same fit indices: Comparative Fit Index (CFI \textgreater 0.95), Tucker Lewis Index (TLI \textgreater 0.95), Root Mean Square Error of Approximation (RMSEA \textless 0.06), Standardized Root Mean Square Residual (SRMR \textless 0.08) and low Chi-square \cite{hu1999cutoff,hinkin1998brief,boateng2018best}. Only SRMR (0.068) passed the threshold, (CFI = 0.836, TLI = 0.811, RMSEA = 0.138), making the condition (1) not hold. Table \ref{table:factor} shown that both conditions (2 and 3) hold, with average factor loading is above .7 and the differences between individual factor loadings and average factor loading are less than .2. Further test and evaluation should be done to confirm this. Nonetheless, we conclude that our measurement instrument is reliable based on the coefficient alpha and omegas. 

\begin{table}[]
\centering

\begin{tabular}{llr}
\hline
\textbf{Parameter} &  & \multicolumn{1}{l}{\textbf{Difference}} \\\hline
Average loading & 0.737 & \multicolumn{1}{l}{} \\\hline
Individual loading &  & \multicolumn{1}{l}{} \\
understandability1 & 0.567 & 0.17 \\
understandability2 & 0.623 & 0.114 \\
technical competence1 & 0.753 & -0.016 \\
technical competence2 & 0.794 & -0.057 \\
reliability1 & 0.759 & -0.022 \\
reliability2 & 0.663 & 0.074 \\
helpful1 & 0.762 & -0.025 \\
helpful2 & 0.824 & -0.087 \\
personal attachment1 & 0.697 & 0.04 \\
personal attachment2 & 0.819 & -0.082 \\
user autonomy1 & 0.740 & -0.003 \\
user autonomy2 & 0.677 & 0.06 \\
faith1 & 0.691 & 0.046 \\
faith2 & 0.780 & -0.043 \\
institution credibility1 & 0.812 & -0.075 \\
institution credibility2 & 0.838 & -0.101 \\\hline
\end{tabular}
\caption{Average loading, Individual item loadings, and Difference between the two, in Fitted Unidimensional (Single Factor) Model.}
\label{table:factor}
\end{table}

\subsubsection{Test-retest reliability from Repeated Survey}
The Mechanical Turk tasks were up for one month and data from 304 participants were collected. The participants were 53.6\% male, 45.4\% female, 0.7\% prefer not to say their gender. Half of the participants were between 30-40 years old (52.6\%), with the rest 19.4\% were between 20-30 years old, 15.8\%  were between 40-50 years old, and 12.2\% above 50 years old. As mentioned previously, the test was quantified using Intra-class Correlation Coefficient (ICC) between the ratings given by the same participants answered at closely spaced points in time (30 minutes - one hour). The test-retest reliability was established with ICC value = 0.7377. 

\section{Test of Validity}
\subsection{Methods}
A measurement instrument should not only be reliable but also valid, because reliability is a necessary but not sufficient condition for validity. Validity refers to the degree to which an instrument accurately measures the dimension or construct for which it was designed \cite{raykov2011introduction}. Due to the wide range of validity assessment metrics, many terms are used, such as, concurrent validity, construct validity, content validity, convergent validity, criterion validity, discriminant validity, divergent validity, face validity, and predictive validity. However, validity assessment can be summarised in three main forms: 
\begin{itemize}
\item Content validity (including: face validity)
\item Construct validity (including: convergent validity, discriminant validity)
\item Criterion validity (including: predictive validity, concurrent validity)
\end{itemize}

Up until this stage, we have evaluated the content validity and face validity in the Item Evaluation stage. Content validity is the degree to which the instrument measure what it designed to measure. Content validity is not only assessed using statistical procedures but also relies on the reasoning \cite{nunnally1994psychometric}, which is why Expert Validation and Cognitive Interview were conducted. 

\subsubsection*{\textbf{Construct Validity}}
After survey has been administered, construct validity of a measurement instrument can be examined. Construct validity refers to the degree to which an instrument assesses a construct of real-world concern and is associated with evidence that measures other constructs  \cite{raykov2011introduction}. Convergent validity is the extent to which a construct measured in different ways produces similar results. Literature suggest that convergent validity is established when the average variance extracted (AVE) is above 0.5 \cite{chin2010write,henseler2009use,bagozzi1988evaluation}. Other literature suggest that alongside AVE value, composite reliability (CR)  also need to be considered, and convergent validity is established when the CR is above 0.7 \cite{hair2019development}.

There are various definitions of discriminant validity in the existing studies, one is the extent to which a measure is novel and not just a reflection of some other measurement \cite{campbell1959convergent,raykov2011introduction,boateng2018best}. The other is the extent to which two constructs were empirically distinguishable \cite{mata2020development}\cite{hu2015making}. Since our main objective is to evaluate the validity of our measurement instrument, we assessed discriminant validity through our trust factors (domains) and trust propensity. However, we also looked at the discriminant validity between our domains to see if our domains are empirically different one from another. 
To evaluate discriminant validity, Fornell-Larcker criterion and Heterotrait-monotrait (HTMT) ratio of correlation can be examined. Fornell-Lacker criterion compares the square root of the AVE with the correlation of latent constructs, and a greater value of each AVE indicates discriminant validity \cite{fornell1981evaluating}. The Heterotrait-monotrait (HTMT) criterion calculates HTMT ratio, and value close to 1 indicates a lack of discriminant validity \cite{henseler2015new}. Literature proposed threshold values for HTMT ratio are 0.9 \cite{gold2001knowledge} and 0.85 \cite{voorhees2016discriminant}. We examined both methods, since it is recommended to employ more than one method \cite{campbell1959convergent}, and literature on discriminant validation have used multiple distinct measurement methods \cite{le2009multifaceted,woehr2012examination} .

\subsubsection*{\textbf{Criterion Validity}}
The content and construct validity are considered as validity of measurement\cite{murphy1988psychological}, and included in the objective tests for psychological instrument \cite{loevinger1957objective}. In contrast, criterion validity is recognised as validity for decision, instead of validity of a measurement \cite{murphy1988psychological}. Validity for decision means that the measurement can lead to a correct/right decision, such as, psychological test for recruitment criteria. Criterion validity refers to the extent of the relationship between a particular criteria on other related measures, whether it's with the "gold-standard" measurement (concurrent validity) or other related measurement in the future (predictive validity) \cite{raykov2011introduction}. Even though our trust measurement was not designed to be a decision-making aid nor designed to predict any future decision, we still assessed the criterion validity of our measurement to explore our measurement aptitude. The criteria we used was trust, and since currently there is no "gold-standard" for human-AI trust measurement, the concurrent validity was assessed with single-question trust measurement from Replication survey. Concurrent validity is established when the constructs of our measurement instrument are significantly correlated to the trust rating, which were measured at the same time. Predictive validity is established by regressing some future outcome on the established construct, which we assessed with regression analysis to the single-question trust measurement from Replication survey.

\subsection{Results}
We were looking at the construct validity, where we evaluate if an instrument measures a construct that is not directly observable \cite{nunnally1994psychometric}. Construct validity composed of convergent validity, when the domains address the same construct; and discriminant validity, when the domains address different aspects of the construct.

\subsubsection*{\textbf{Construct Validity: Convergent Validity}}
A measurement can established convergent validity when the measurement domain/construct correlates highly with each other. As depicts in Table \ref{table:crave}, all composite reliability (CR) values are above 0.7 \cite{hair2019development}, all average variance extracted (AVE) are above 0.5, suggesting the convergent validity of measurement instrument. This result suggest that our measurement instrument could examines or measures trust in different ways while still yields similar results.

\begin{table}[]
\makebox[\textwidth]{
\begin{tabular}{|l|l|l|llllllll|}
\hline
 & \multicolumn{1}{c|}{CR} & \multicolumn{1}{c|}{AVE} & \multicolumn{1}{c}{r} & \multicolumn{1}{c}{tc} & \multicolumn{1}{c}{u} & \multicolumn{1}{c}{pa} & \multicolumn{1}{c}{h} & \multicolumn{1}{c}{f} & \multicolumn{1}{c}{ua} & \multicolumn{1}{c|}{ic} \\\hline
reliability (r) & 0.724 & 0.585 & \textbf{0.764} &  &  &  &  &  &  &  \\
technical competence (tc) & 0.838 & 0.733 & 0.895* & \textbf{0.855} &  &  &  &  &  &  \\
understandability (u) & 0.794 & 0.665 & 0.772* & 0.815 & \textbf{0.815} &  &  &  &  &  \\
personal attachment (pa) & 0.839 & 0.729 & 0.795* & 0.748 & 0.552 & \textbf{0.853} &  &  &  &  \\
helpfulness (h) & 0.870 & 0.769 & 0.939* & 0.870* & 0.747 & 0.746 & \textbf{0.876} &  &  &  \\
faith (f) & 0.831 & 0.708 & 0.803* & 0.683 & 0.479 & 0.833 & 0.729 & \textbf{0.841} &  &  \\
user autonomy (ua) & 0.831 & 0.713 & 0.758* & 0.764 & 0.639 & 0.751 & 0.731 & 0.694 & \textbf{0.844} &  \\
institution credibility (ic) & 0.914 & 0.846 & 0.779* & 0.774 & 0.552 & 0.836 & 0.744 & 0.870* & 0.743 & \textbf{0.920} \\ \hline
\end{tabular}}
\caption{Composite reliability (CR), the average variance extracted (AVE), the square root of AVE (in bold), and correlations between constructs (off-diagonal). }
\label{table:crave}
\end{table}

\subsubsection*{\textbf{Construct Validity: Discriminant Validity}}
The discriminant validity of inter-construct was first assessed with Fornell and Larcker criterion \cite{fornell1981evaluating}. As described above, this method compares the AVE with the correlation of latent constructs, and a greater value of each AVE indicates discriminant validity. However, there are different interpretation in the literature to conclude the discriminant validity. Some literature applied Henseler et al. \cite{henseler2015new} interpretation, where the comparison is made from the square root of each AVE with the correlation coefficients for each construct \cite{ab2017discriminant}\cite{mata2020development}. Based on this interpretation, as shown in the Table \ref{table:crave}, the correlation coefficients between reliability - all constructs, helpfulness-technical competence, and institution credibility-faith, are greater than the AVEs (marked with *) which means those constructs do not hold discriminant validity. 

Another interpretation claimed that discriminant validity is only established when the AVEs for both constructs are bigger than the squared factor correlation/shared variance (SV) between them, and Henseler et al. interpretation is considered one of the misapplications of Fornell-Larcker criterion \cite{ronkko2022updated}. The misapplication rooted in the usage of only one of the two AVE values or the average of the two AVE values, instead of both AVE values when comparing to the SV. Based on this original interpretation, as shown in the Table \ref{table:avesv}, the SVs between reliability-technical competence, reliability-personal attachment, reliability-helpfulness, reliability-faith, helpfulness-technical competence, faith-personal attachment, and institution credibility-faith, are greater than both constructs AVEs (marked with *) which means those constructs do not hold discriminant validity \cite{ronkko2022updated}. 

\begin{table}[]
\makebox[\textwidth]{
\begin{tabular}{|l|llllllll|}
\hline
AVE/SV & \multicolumn{1}{c}{r} & \multicolumn{1}{c}{tc} & \multicolumn{1}{c}{u} & \multicolumn{1}{c}{pa} & \multicolumn{1}{c}{h} & \multicolumn{1}{c}{f} & \multicolumn{1}{c}{ua} & \multicolumn{1}{c|}{ic} \\ \hline
reliability (r) & \textbf{0.585} &  &  &  &  &  &  &  \\
technical competence (tc) & 0.755* & \textbf{0.733} &  &  &  &  &  &  \\
understandability (u) & 0.568 & 0.642 & \textbf{0.665} &  &  &  &  &  \\
personal attachment (pa) & 0.604* & 0.547 & 0.287 & \textbf{0.729} &  &  &  &  \\
helpfulness (h) & 0.854* & 0.745* & 0.543 & 0.545 & \textbf{0.769} &  &  &  \\
faith (f) & 0.627* & 0.467 & 0.228 & 0.759* & 0.523 & \textbf{0.708} &  &  \\
user autonomy (ua) & 0.556 & 0.569 & 0.403 & 0.575 & 0.530 & 0.491 & \textbf{0.713} &  \\
institution credibility (ic) & 0.576 & 0.579 & 0.295 & 0.690 & 0.543 & 0.748* & 0.537 & \textbf{0.846} \\ \hline
trust propensity & 0.180 & 0.203 & 0.133 & 0.374 & 0.163 & 0.433 & 0.242 & 0.328 \\  \hline
\end{tabular}}
\caption{Fornell-Larcker Table: The average variance extracted AVE (in bold), the square root of correlations between constructs SV (off-diagonal).}
\label{table:avesv}
\end{table}

Lastly, we examined discriminant validity with the Heterotrait-monotrait (HTMT) criterion. HTMT criterion was proposed as a superior method that successfully achieved higher specificity and sensitivity rates (97-99\%) compared to the older criterion, including Fornell-Lacker (20.82\%). As mentioned previously, lack of discriminant validity usually indicated with HTMT value close to 1, or HTMT higher than set threshold. Table \ref{table:htmt} shows HTMT results, and using both 0.85 \cite{voorhees2016discriminant}  and 0.9 \cite{gold2001knowledge} threshold, reliability-technical competence, reliability-helpfulness, helpfulness-technical competence, faith-personal attachment, and institution credibility-faith passed the threshold (marked with *). Based on this criterion, discriminant validity was not established on all items. In summary, using different methods and criterion, the discriminant validity was not established on all inter-constructs (domains) of our measurement instrument. Since the dimensionality and the internal consistency of our measurement has been established, similarity between dimensions can be examined further in future research, specifically on trust factors model. 

To evaluate the discriminant validity of our trust measurement, we assessed the relation between our measurement and participants' trust propensity level. As we can see in the last row of Table \ref{table:avesv} and Table \ref{table:htmt}, based on Fornell and Larcker criterion and HTMT ratio, the results suggested low similarity between our trust factors and trust propensity. These results supported the discriminant validity of our measurement instrument, which was different from trust propensity.

\begin{table}[]
\begin{tabular}{|l|llllllll|}
\hline
 & r & tc & u & pa & h & f & ua & ic \\\hline
reliability (r) & 1.000 &  &  &  &  &  &  &  \\
technical competence (tc) & 0.890* & 1.000 &  &  &  &  &  &  \\
understandability (u) & 0.778 & 0.823 & 1.000 &  &  &  &  &  \\
personal attachment (pa) & 0.783 & 0.745 & 0.541 & 1.000 &  &  &  &  \\
helpfulness (h) & 0.945* & 0.873* & 0.742 & 0.737 & 1.000 &  &  &  \\
faith (f) & 0.799 & 0.689 & 0.484 & 0.872* & 0.723 & 1.000 &  &  \\
user autonomy (ua) & 0.765 & 0.765 & 0.645 & 0.762 & 0.729 & 0.705 & 1.000 &  \\
institution credibility (ic) & 0.769 & 0.772 & 0.550 & 0.832 & 0.740 & 0.868* & 0.736 & 1.000\\\hline
trust propensity & 0.425 & 0.455 & 0.368 & 0.612 & 0.404 & 0.658 & 0.492 & 0.573  \\ \hline
\end{tabular}
\caption{The Heterotrait-Monotrait (HTMT) Ratio Table. }
\label{table:htmt}
\end{table}

\subsubsection*{\textbf{Criterion Validity: Concurrent Validity}}
To established concurrent validity, our measurement should be significantly correlated to some outcome measured at the same time, meaning the trust factors should be correlated to the trust level. As shown in Table \ref{table:concurv}, each domain (trust factor) of our measurement is positively correlated to trust level and the relationships are all significant (p < .05). These results suggested that our trust measurement hold concurrent validity towards subjective single-question trust measurement. 

\begin{table}[]
\begin{tabular}{l|llll}
\hline
Domains & Correlation & p-value & lower CI & upper CI \\\hline
reliability & 0.786 & 0.000 & 0.708 & 0.857 \\
technical competence & 0.829 & 0.000 & 0.764 & 0.895 \\
understandability & 0.581 & 0.000 & 0.438 & 0.722 \\
personal attachment & 0.814 & 0.000 & 0.756 & 0.867 \\
helpfulness & 0.813 & 0.000 & 0.755 & 0.871 \\
faith & 0.846 & 0.000 & 0.797 & 0.890 \\
user autonomy & 0.782 & 0.000 & 0.708 & 0.855 \\
institution credibility & 0.779 & 0.000 & 0.713 & 0.841\\\hline
\end{tabular}
\caption{Correlation Table: Domains (Trust Factors) with Trust. Correlation Coefficient, p-value, lower Confidence Interval, upper Confidence Interval. }
\label{table:concurv}
\end{table}

\subsubsection*{\textbf{Criterion Validity: Predictive Validity}}
For predictive validity, first, we regressed trust level data on all items rating, with linear regression model: trust equal all trust factors items. As shown in Table \ref{table:predv}, two faith items and one institution credibility item were significantly predictive of trust level (p \(<\) .05). Based on these results, the relationships between trust and all trust factors, except Faith, were not linear. Thus, trust level could not be predicted with 16 items and the predictive validity of our measurement was not supported. 

\begin{table}[]
\begin{tabular}{l|lllll}
\hline
Items & Reg Coeff & p-value &  lower CI & upper CI & \(R^2\) \\\hline
reliability1 & -0.014 & 0.813 & -0.128 & 0.100 & 0.713 \\
reliability2 & 0.057 & 0.259 & -0.042 & 0.155 & 0.612 \\
technicalcompetence1 & 0.113 & 0.057 & -0.004 & 0.229 & 0.656 \\
technicalcompetence2 & 0.113 & 0.106 & -0.024 & 0.250 & 0.703 \\
understandability1 & -0.027 & 0.638 & -0.137 & 0.084 & 0.698 \\
understandability2 & 0.016 & 0.764 & -0.090 & 0.123 & 0.652 \\
personalattachment1 & 0.074 & 0.157 & -0.029 & 0.177 & 0.698 \\
personalattachment2 & 0.024 & 0.672 & -0.087 & 0.135 & 0.808 \\
helpful1 & 0.027 & 0.671 & -0.096 & 0.149 & 0.747 \\
helpful2 & 0.092 & 0.154 & -0.035 & 0.220 & 0.797 \\
\textbf{faith1} & \textbf{0.230} & \textbf{0.000} & \textbf{0.109} & \textbf{0.351} & \textbf{0.791} \\
\textbf{faith2} & \textbf{0.214} & \textbf{0.001} & \textbf{0.085} & \textbf{0.342} & \textbf{0.830} \\
userautonomy1 & -0.069 & 0.223 & -0.181 & 0.042 & 0.705 \\
userautonomy2 & 0.078 & 0.142 & -0.026 & 0.182 & 0.702 \\
\textbf{institutioncredibility1} & \textbf{0.167} & \textbf{0.005} & \textbf{0.050} & \textbf{0.285} & \textbf{0.800} \\
institutioncredibility2 & -0.118 & 0.084 & -0.251 & 0.016 & 0.841 \\\hline
\end{tabular}
\caption{Regression Table: Measurement Items with Trust. Regression Coefficient, p-value, lower Confidence Interval, upper Confidence Interval, and Coefficient of Determination \((R^2)\) }
\label{table:predv}
\end{table}

\section{Discussion}
The main purpose of this study was to develop and validate a scale to measure the trust attitude of lay people towards AI systems. Eight domain factors and 16 items were developed through deductive and inductive methods and have been thoroughly evaluated. The proposed measurement instrument was administered as an online survey. The results have shown that the measurement instrument is internally consistent and stable; with established content and construct validity.

Guidelines from APA \cite{american1999standards} and literature \cite{boateng2018best,hinkin1998brief} have emphasised the importance of content validity, criterion-related validity, and internal consistency of measurement instruments. However, to date, not all studies that use psychological construct instruments have demonstrated both their validity and reliability. The internal consistency of our instrument was established for all measured domains with alpha and omega coefficients. In recent years, research has encouraged the use of the omega coefficient and considers it a better choice than the alpha coefficient for assessing \cite{cho2015cronbach} reliability. However, the alpha coefficient is still a popular choice among trust measurement instruments \cite{madsen2000measuring,mcknight1998initial,jian2000foundations}. A possible explanation is that alpha is considered more familiar than omega, and the difference between alpha and omega is also believed to be small \cite{deng2017testing}.

In addition to internal consistency, stability over time was assessed by test-retest reliability and resulted in a high correlation (ICC) for the 1-hour time between survey administrations. It should be noted that test-retest reliability requires a short duration between administrations to allow changes to occur, but still long enough to prevent fatigue and preserve memory \cite{schultz2005measurement}. 
We repeated the survey with a different group of participants, as it is advisable to use different samples \cite{campbell1959convergent} while helping to increase the generalisability of the measurement \cite{hinkin1998brief}. Stability over time indicates repeatability of the measurement, and a measurement cannot be valid if it cannot be repeated.

Although reliability is necessary and should be clearly reported for all measures in a study, it is not a sufficient condition for measurement validity. Studies report that most research on trust in the Human Interaction-AI uses readily available trust questionnaires derived from Jian et al. \cite{lee2004trust}, Chien et al.\cite{chien2018effect}, Merritt \cite{merritt2011affective}, and Muir \cite{muir1987trust}; which are often not assessed for more than internal consistency reliability \cite{spain2008towards}. 
In addition, half of the human-AI trust papers modified the original questionnaire intended for a different system (automation) without providing any validation tests. 

A measurement is considered valid when it measures what it is intended to measure, with two main types of validity: content validity and construct validity \cite{murphy1988psychological}. Content validity refers to evidence of the representativeness of the content and technical quality of the instrument items, which in our study was assessed quantitatively using the content validity index (CVI) and Cohen's kappa coefficient (\(\kappa\)), and qualitatively using cognitive interviews. Most popular studies have failed to clearly indicate the content validity assessment of their questionnaires \cite{jian2000foundations,chien2018effect,merritt2011affective}, which can cause problems in studies that use questionnaires even without changes or modifications. We further validated the measurement qualitatively with cognitive interviews. For questionnaires aimed at specific groups of people, it is very important to check whether the designed instrument is interpreted correctly. In the cognitive interview, we found that participants were able to describe their mental processes when answering questions from our measurement instrument. As none of the participants had experience in AI medical support systems, their reflection processes were related to more common AI applications, such as, Google Maps, Google Translate, Instagram Recommendations, etc. This implies that the measurement instrument was designed for a specific group of people. This implies that our trust measurement instrument might be used to evaluate AI systems in general and not limited to AI medical support systems. This implication may be somewhat limited as a study has shown that trust in technology and trust in medical technology have the same attributes but are considered to be different constructs \cite{montague2009empirically}.

The domains included in our measurement were also tested to check if our hypothetical structure of the eight factors model fits the items. The dimensions are perceived reliability, perceived technical competence, perceived understandability, faith, personal attachment, perceived helpfulness, user autonomy, and institution credibility; with two items on each dimension. Confirmatory factor analysis (CFA) was performed, and the results have shown that the structural validity of the instrument was established. Based on the CFA model, construct validity was analysed, which refers to the evidence of instrument capability to measure a construct that is not directly observable. In our study, we assessed the construct validity: convergent validity and discriminant validity, of our measurement instrument using Fornell-Larcker's criterion \cite{fornell1981evaluating}. Convergent validity was established between all domain items, with all composite reliability (CR) values and AVE values passing the minimum threshold, proving that all domains explain a substantial amount of variance in each indicator. As mentioned in the previous section, the Fornell-Larcker criterion requires the AVE to be greater than 0.5, and research by Hair et al. suggested that not only AVE should be greater than 0.5 but CR should also be greater than 0.7 \cite{hair2019development}. 

Discriminant validity was tested for the overall measurement and also for between domains. Based on Fornell-Larcker's and HTMT ratio criterion, the discriminant validity between trust propensity and our trust measurement was established. This finding supports the work of other studies in trust theory that distinctly defined trust and trust propensity as conceptually different \cite{mayer1995integrative,rousseau1998not,mcknight2011trust,colquitt2007trust}. In contrast, discriminant validity between all domains was not established. Based on Fornell-Larcker's and HTMT ratio criterion, six out of eight domains were not demonstrating discriminant validity: reliability-technical competence, reliability-helpfulness, helpfulness-technical competence, faith-personal attachment, and institution credibility-faith. The high correlation between domains indicates a potential problem in some of the domain differentiation. Even though discriminant validity is a matter of degree instead of binary \cite{ronkko2022updated}, and only five out of 28 correlations demonstrated low discriminant validity, the possible cause should be analysed further. There is a high chance for factors affecting trust, affect each other as well. 

Criterion-related validity was assessed for both concurrent validity and predictive validity. The "criteria" we used was trust level, and concurrent validity was established. According to Raykov \& Marcoulides \cite{raykov2011introduction}, concurrent validity is often omitted from validation study because it has two major constraints which are: the availability of appropriate criterion variables or "gold-standards" and the large sampling errors for small sample size. Even though our sample size was adequate, the assessment was ran on the assumption that single-question trust level is the gold-standard of trust measurement. This proposed a question for future studies on how to appropriately assess concurrent validity on a trust measurement instrument. It is argued that trust is context heavy, where different researchers could defines trust in a widely different way, and currently, a widely accepted and used measurement instrument for trust 
from an empirical standpoint \cite{watson2005can} has not emerged to be constituted as the gold standard. Another important point to note, concurrent validity is a part of criterion validity, which is the validity of decision made by measurement. Taken together, a reasonable approach to assess concurrent validity of a trust measurement instrument is with trust-related behaviour (decision) measurement, such as reliance. Same with predictive validity, a more appropriate approach to assess predictive validity of a trust measurement instrument is by means of a trust-related behaviour (decision) measurement. When a trust measurement instrument established its predictive validity, it means the measurement scores can determine future outcomes. Even though predictive validity was not demonstrated for our measurement instrument to predict trust level, this did not diminish the other types of validity of our measurement instrument. There is also the possibility of a non-linear relation between trust factors and trust level, and further tests need to be conducted to explore these relations better. 

To analysed the content validity, convergent validity, discriminant validity, concurrent validity, and predictive validity, different tests and assessments were performed. As described in the result subsections for each validity test, there are discussions in psychometric field on the better method and thresholds recommended, such as HTMT ratio criterion as alternative method from Fornell-Lercker criterion criticism \cite{henseler2015new}. Another example of discussion was made to both HTMT and Fornell-Lercker criterion, wich were argued to be ineffective in evaluating convergent validity and discriminant validity \cite{cheung2017current}. However, applying Cheung and Wang \cite{cheung2017current} recommendation to conclude convergent validity (AVE is not significantly smaller than 0.5 and the standardised item factor loading item is not significantly less than 0) and discriminant validity (correlation less than 0.7), the previous results of our measurement instrument still hold. Criticism on methods is outside of our research scope, therefore, we applied various methods to empirically test our measurement instrument. The results from all validity and reliability tests increased the confidence that the instrument correctly measured the construct intended to measure.

\section{Limitations}
Several limitations of the study and possible future works can be highlighted on the three main stages: item development, measurement instrument development, and measurement instrument evaluation. In the item development stage, we looked at the literature from different fields, not only on trust scales but also on trust factors and models. Even though we proposed eight factors that could affect trust to form our trust measurement instrument, we have not analysed closely the relation between factors, and no actual trust model for human-AI system interaction has been proposed. Future analysis and, if required, data collection should be conducted to develop the trust model. In the measurement instrument development stage, we did not include detailed demography questions in the survey administered.  Additional demography questions could help understand trust based on the demographics better, and also could help development of the trust model. 

Repeatability is an important part of a measurement instrument, and we conducted an additional survey to assess the test-retest reliability of our measurement. Since theory suggests that trust is dynamic and may vary over time, future longitudinal research utilising our measurement instrument would be appropriate to examine the test-retest reliability for a longer time period. Instead of just an hour, repeated survey with 1-7 days in between is recommended. Additionally, our repeated survey was conducted with the same topic: breast cancer detection applications. Replication with different types of AI systems, possibly outside of healthcare application, would be beneficial to developing the trust model and also solidify the generality of our measurement instrument. In the discussion section previously, we mentioned that further evaluation will be necessary to understand trust and develop the trust theory/model. A future study that involves the measurement of trust as an attitude and how it relates to trust-related behaviour could also provide a valuable contribution. 

In summary, replication and adaptation of our proposed trust measurement instrument are highly encouraged. The trust scale replication and validation by Spain et al. \cite{spain2008towards} was one of the reasons why Jian et al. \cite{jian2000foundations} trust scale became the most well-cited trust scale in the human factors literature \cite{gutzwiller2019positive}. Replication and adaptation will not only to further prove the validity and generality of the measurement but also will help to understand trust and trust model in human-AI system interaction. 

\section{Conclusion}
Trust is an important concept, and trust in Artificial Intelligence is currently widely explored in various research fields. This study makes theoretical and practical contributions by developing and validating a scale to measure trust attitudes towards AI systems for laypeople. We proposed a trust measurement comprised of eight trust factors as the domains and two statement items for each domain, which make a total of 16 items. The reliability and validity of our measurement instrument were established and are expected to be used and adapted by future researchers to evaluate their AI systems. 

The methodological approach to develop and evaluate trust measurement for human-AI interaction has been described and demonstrated. Carefully designed trust measurement instruments are not only fundamental to our understanding of trust but also ensure accurate measurement of trust, which is known to be a complex construct. By making the development of measurement instruments more approachable and transparent, we hope this paper can facilitate the advancement of our understanding of trust and trustworthiness in AI systems while complementing existing trust models.







\end{document}